\documentclass[aps,prb,twocolumn,showpacs,amsmath,amssymb]{revtex4}
\usepackage{dcolumn}
\usepackage{bm}
\usepackage{graphicx}
\usepackage{bm}
\begin{document}
\title{Kinetic Arrest of Field-Temperature Induced First Order Phase Transition in Quasi- One Dimensional Spin System  Ca$ {_3} $Co$ {_2} $O$ {_6} $}
\author{Santanu De, Kranti Kumar, A. Banerjee and P. Chaddah}
\affiliation{UGC-DAE Consortium for Scientific Research, University Campus, Khandwa Road, Indore-452001, India.}

\begin{abstract}
We have found that the geometrically frustrated spin chain compound Ca$ {_3} $Co$ {_2} $O$ {_6} $ belonging to Ising like universality class with uniaxial anisotropy shows kinetic arrest of first order intermediate phase (IP) to ferrimagnetic (FIM) transition. In this system, dc magnetization measurements followed by different protocols suggest the coexistence of high temperature IP with equilibrium FIM phase in low temperature. Formation of metastable state due to hindered first order transition has also been probed through cooling and heating in unequal field (CHUF) protocol. Kinetically arrested high temperature IP appears to persist down to almost the spin freezing temperature in this system.
\end{abstract}
\maketitle
\section {Introduction}
Kinetic arrest of magnetic first order phase transition (FOPT) is defined as the arrest of kinetic energy of the molecules in a thermo-magnetic system before the extraction of latent heat which involve with the FOPT of that system. It has been studied widely during the last decade in different class of materials like half doped perovskite manganites, intermetallics and shape memory alloys. \cite{1,2,3,4,5,6}
Among the spin chain systems a class of material with general formula  A$^{\prime} {_3} $ABO$ {_6} $ (where A$^{\prime}$ can be Ca, Sr, Ba and A, B are respectively metallic and magnetic or nonmagnetic ions) has drawn appreciable attention due to their exotic magnetic properties. \cite{rev,8,10,11,12,14,15,16,17,18,20} Presence of competing interactions in these geometrically frustrated systems gives rise to interesting magnetism. The compound Ca$ {_3} $Co$ {_2} $O$ {_6} $ crystallizes in rhombohedral crystal system having spin chains made up of CoO$ {_6} $ trigonal prisms (TP) and CoO$ {_6} $ octahedra (OCT) sites along c-axis of the hexagonal crystal structure. \cite{rev} Each chain of spins has six neighbouring chains and two consecutive chains are separated by calcium ion. Along the chain there is a strong ferromagnetic coupling between Co$^{3+}$ ions with S = 2, whereas in ab plane coupling is weakly antiferromagnetic in a triangular sub-lattice. \cite{rev,8} At low temperature (low-T) both intrachain and interchain coupling becomes strong enough as a result the quasi 1-D spin system tends to higher dimensions and confirms the possibility of  long range order (LRO). A brief review on this compound includes the following outcomes which will be useful for our discussions: 

(1)  There is a controversial transition from paramagnetic to an intermediate phase (IP) around T$_{I}$  $\sim $  26 K in zero field, initially it was reported as partially disordered antiferromagnetic \cite{10,11,12,20} (where 2/3 of chains are antiferromagnetically coupled  and rest of 1/3  left disordered in a triangular sub-lattice) whereas recent reports suggest the spin density wave ordering. \cite{18,17,16}
(2) Below T$_{I}$ in small nonzero magnetic field, the possible magnetic state is ferrimagnetic (FIM) and with increase in field it progress towards the ferromagnetic state (FM).  \cite{rev,8,10,11,12,18,20}

In this paper we investigate the nature of the transition from IP to FIM in the spin chain system Ca$ {_3}$Co${_2}$O${_6}$. We find that the IP to FIM transition is a broad first order transition. However, there is magnetic field and temperature induced hindrance to the transformation kinetics which masks the first order IP to FIM transition. As a result there is coexistence of equilibrium FIM phase with a fraction of high temperature (high-T) IP in low-T. Metastable phase fraction depends on the path traced in H-T plane. Temperature dependent dc magnetization in a magnetic field shows a thermal hysteresis in the range of T $\approx$ 8-20 K. M-H curves at 10 K depicts that the virgin path is not followed in subsequent field cycling. Moreover, the measurement of M-T in cooling and heating in unequal field (CHUF) protocol \cite{3} argue that larger fraction of glass-like arrested IP exists in lower field down to about 6 K. It also says that this arrested metastable IP devitrifies to FIM phase while warming when the measuring field (H$_{m}$) $>$ cooling field (H$_{c}$).

\maketitle\section{Experimental Details}
Polycrystalline Ca$ {_3} $Co$ {_2} $O$ {_6} $ was synthesized through solid state route. Details of sample preparation method are given in the references [16 and 17]. All the measurements were carried out using 14 T vibrating sample magnetometer (VSM) with temperature range 2 - 350 K and low field measurement were performed in superconducting quantum interference device (SQUID) magnetometer with applied magnetic field ranging from -7 T to +7 T and temperature down to 2 K. 

\maketitle\section{Results and discussions}
This quasi 1-D compound undergo a paramagnetic to IP transformation around T$_{I}$ $\sim $ 26 K at H = 0  \cite{10,11,12,20,18,17,16} and below that in presence of small applied magnetic field it goes to FIM state at low-T. \cite{rev,8,10,11,12,18,20} Temperature dependent dc magnetization of the compound Ca$ {_3} $Co$ {_2} $O$ {_6} $ in applied magnetic field (H) 0.001 T and 0.01 T is shown in FIG. 1 (a) and (b) respectively using three different protocols. Magnetization under zero field cooled warming (ZFC) and field cooled warming (FCW) mode coincided with each other down to  $\sim $ 9 K and below that they bifurcate. M-T curve exhibits a thermal hysteresis in between field cooled cooling (FCC) and FCW data in both the fields depicting a broad first order transition from IP to FIM phase. The main interesting outcome is the bifurcation between ZFC and FCC/FCW which increases enormously below 9 K. This feature indicates an incomplete broad FOPT in both cases. Unlike metastability arising from spin-glass like phenomena, superparamagnetism or anisotropy driven systems, here the bifurcation increases around 12 times for 0.01 T compared to 0.001 T case. This is one of the evidence of kinetically arrested IP at low-T.\cite{1,2,3,4,5,6}  Arrested phase fraction of IP is larger in lower field than the higher field. Therefore it can be concluded that a broad first order IP to FIM phase transition is occurring which is incomplete in nature. 
\begin{figure}[htbp]
	\centering
		\includegraphics[width=8cm]{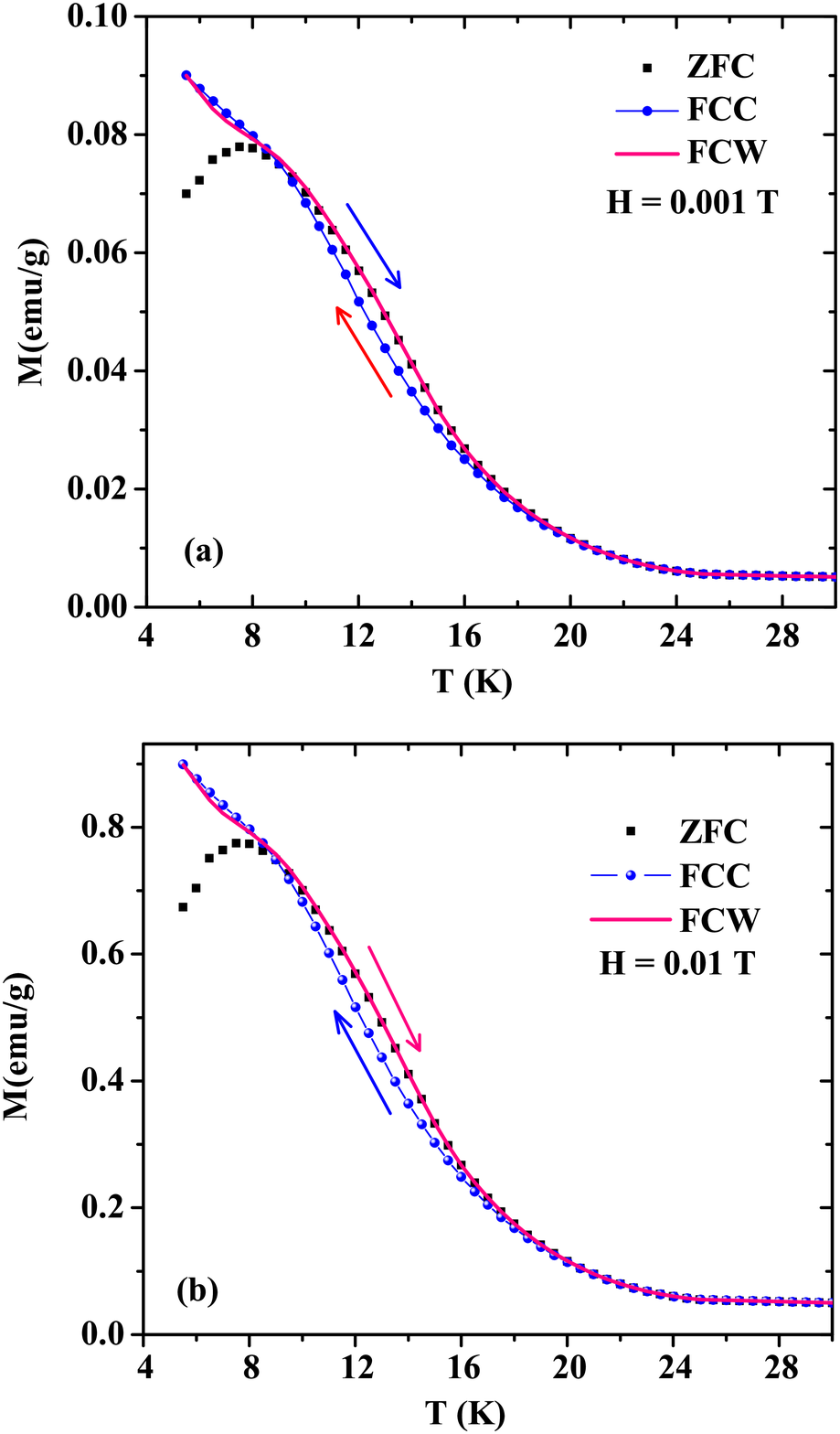}
		\caption{(Color online) Temperature dependent dc magnetization in applied magnetic field (a) H = 0.001 T and (b) H = 0.01 T using ZFC, FCC, FCW protocols.}
\label{fig:fig1}
\end{figure}

FIG. 2 (a) illustrates the field dependence of isothermal (at 10 K) dc magnetization  which revealed that zero field cooled IP converted to FIM phase in application of magnetic field and remains in that state upto 3.4 T  with M $\sim $ 14 emu/g in our case. Whereas earlier reported upper field limit for this state is 3.6 T in single crystal. Further increase in field allows the system to approach towards the FM state. \cite{rev,8,10,11,12,18,20} FIG. 2 (b) shows the expanded view of first step. Field decreasing branch of it from 2.4 to 0 T (whereas M-H measurement was performed in the complete field range of -14 T to +14 T for all the cycles) does not return to high-T IP and subsequent field increasing branch emphasizes the presence of entirely FIM phase. As a result envelop of M-H curve at 10 K goes above the virgin curve. Same kind of behaviour was reported in a variety of functional materials. \cite{1,2,3,4,5,6} Such peculiar behaviour of this material is ascribed to hindrance of the FOPT process by critically slowing down the transition with divergence of time scale. This anomaly is quite similar to glass like freezing of the supercooled high-T IP. Therefore this phenomenon is a manifestation of hindered transformation or kinetic arrest of FOPT. \cite{2}
\begin{figure}[htbp]
	\centering
	\includegraphics[width=8cm]{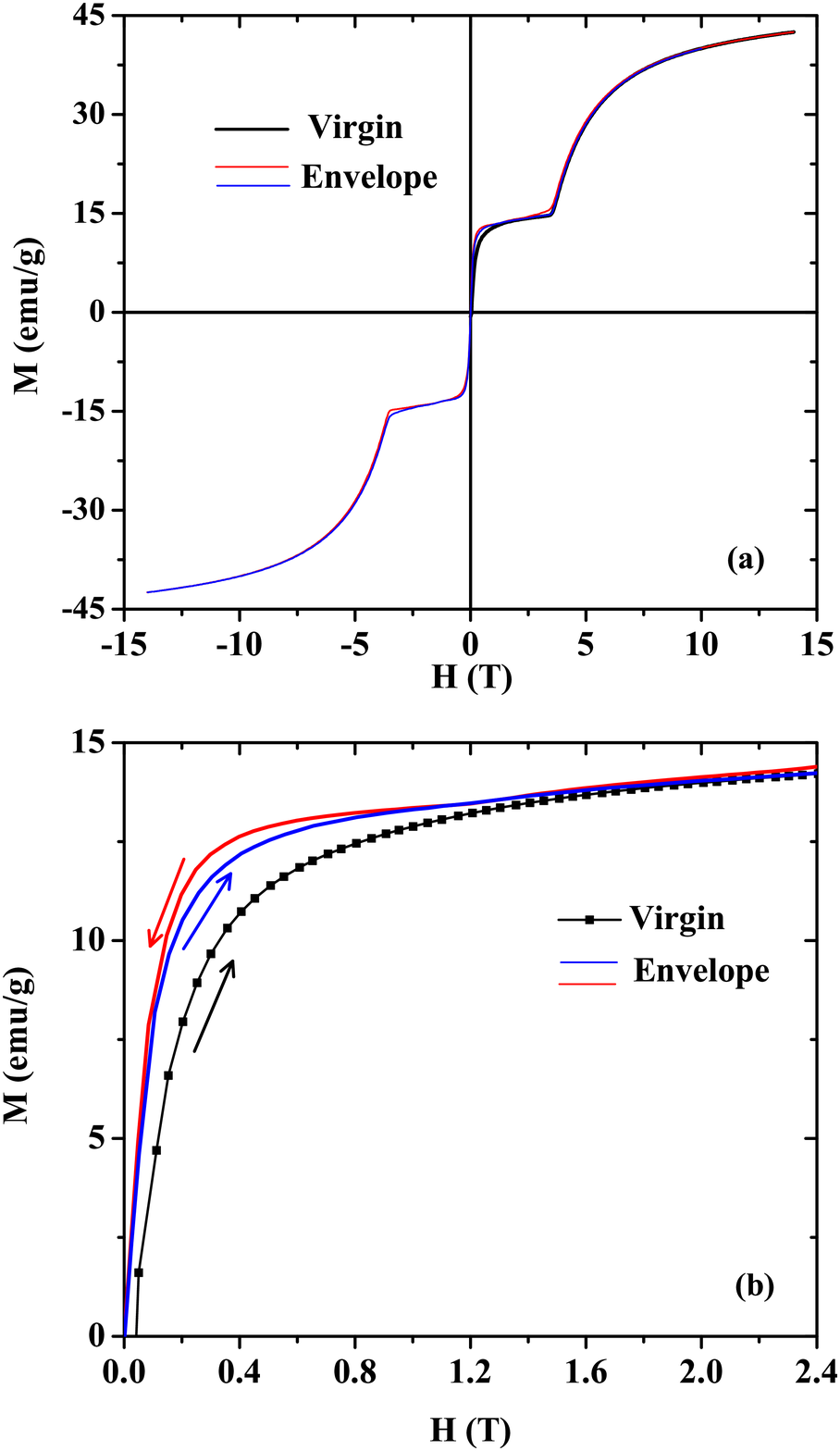}
	\caption{(Color online) (a) Field dependent dc magnetization at 10 K and (b) Displays the zoomed section of FIG. 2 (a) in the field range 0-2.4 T.}
	\label{fig:fig2}
\end{figure}

In CHUF protocol,\cite{3} we cool the sample in zero or nonzero magnetic field and measure the magnetization of the sample during warming cycle in a different magnetic field. It is one of the important tools to probe kinetic arrest of FOPT. The variation of magnetization in CHUF mode is shown in FIG. 3. Cooling in lower field exhibits the arrest of a large percentage of high-T phase (IP). There are two changes of opposite sign in magnetization data when (H$_{m}$ - H$_{c}$) is positive. This corresponds to devitrification of IP which results in rapid increase in magnetization i.e. the system tends to the equilibrium FIM phase followed by a decrease in magnetization corresponding to first order transformation of FIM to IP with increase in temperature. On the contrary, for negative difference between (H$_{m}$) and (H$_{c}$) magnetization decreases monotonically with the increase in temperature  which corresponds to the conversion of equilibrium low-T FIM phase to IP, onsets above superheating-T. \cite{1,2,3,4,5,6}
\begin{figure}
	\centering
	\includegraphics[width=8cm]{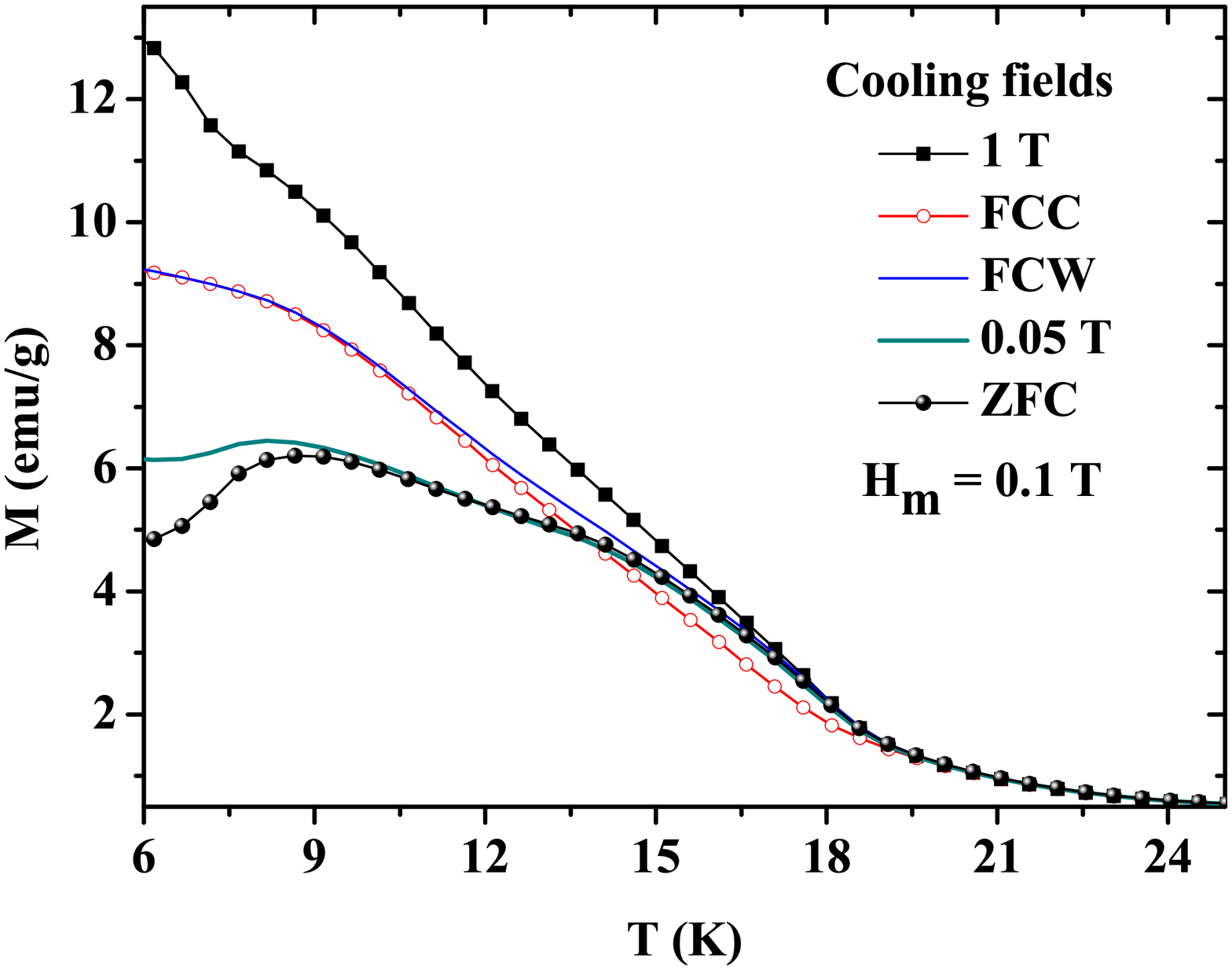}
	\caption{(Color online) M-T data in CHUF mode in the measuring field of 0.1 T. Warming data in 0.1 T after cooling in 1 as well as 0.05 T along with the ZFC, FCC and FCW are depicted.  }

\end{figure}
\maketitle\section{Conclusion}
A new phenomenon is discovered by observing the temperature and field dependence of dc magnetization in this quasi-one dimensional spin system Ca$_{3}$Co$_{2}$O$_{6}$. The system undergo a glass like freezing of high-T IP because the kinetics of molecules of the field induced first order IP to FIM transformation is arrested by critically slowing down the transformation. Competing interactions in this system might have played an important role for critical slowdown of transformation.
\maketitle\section{Acknowledgment}
We are grateful to Professor E. V. Sampathkumaran for providing the sample. SD thanks Mr. Pallab Bag for fruitful discussions.

\end{document}